\documentclass[prl,twocolumn,showpacs,amsmath,amssymb]{revtex4}

\usepackage{graphicx}
\usepackage{dcolumn}
\usepackage{bm}
\usepackage{epsfig}

\begin{document}

\title{Shot Noise Probing of Magnetic Ordering in Zigzag Graphene Nanoribbons}

\author{Ralitsa L. Dragomirova}
\author{Denis A. Areshkin}
\author{Branislav K. Nikoli\' c}

\affiliation{Department of Physics and Astronomy, University of Delaware, Newark, DE 19716-2570, USA}

\begin{abstract}
The nonequilibrium time-dependent fluctuations of charge current have recently emerged as a sensitive experimental tool to probe ballistic transport through evanescent wave functions introduced into clean wide and short graphene strips by the attached metallic electrodes. We demonstrate that such ``pseudo-diffusive'' shot noise can be substantially modified in zigzag graphene nanoribbon (ZGNR) due to the topology of its edges responsible for localized states that facilitate ferromagnetic ordering along the edge when Coulomb interaction is taken into account. Thus, the shot noise enhancement of unpolarized, and even more sensitively of spin-polarized, charge currents injected into ZGNR will act as an {\em all-electrical} and {\em edge-sensitive} probe of such low-dimensional magnetism.
\end{abstract}

\maketitle

The advent of graphene~\cite{Geim2007}---first truly two-dimensional crystal whose carbon atoms form a honeycomb lattice---has reinvigorated exploration of low-dimensional quantum transport phenomena. The high-mobility graphene flakes are far more accessible to different experimental and fabrication techniques than a traditional two-dimensional electron gas (2DEG). Furthermore, the chiral massless Dirac fermions carrying the current in graphene are quite different from the usual quasiparticles in 2DEGs. Thus, many standard mesoscopic transport experiments  have been reexamined in this new 2D setting to unveil their anomalous versions~\cite{Geim2007} due to relativistic-like character of low energy electronic excitations. In particular, very recent experiments on the shot noise in ballistic~\cite{Danneau2008} and disordered~\cite{DiCarlo2008} graphene strips elucidate fundamental conduction properties of Dirac fermions that cannot be extracted from conventional time-averaged current $\bar{I}$.

The shot noise denotes time-dependent current fluctuations,  driven by the nonequilibrium state, which
persist down to zero temperature and originate from the granularity of charge~\cite{Blanter2000}. The zero-frequency noise power $S$
and related Fano factor $F=S/2e\bar{I}$ can probe the effects of disorder, carrier statistics, and interactions in samples
smaller than the electron-phonon inelastic scattering length~\cite{Danneau2008,Blanter2000}. The Poissonian limit $F=1$ characterizes  transport governed by uncorrelated stochastic processes (as encountered in tunnel junctions). More intriguing sub-Poissonian Fano factors are found in, e.g.,  noninteracting diffusive conductors where $F=1/3$ is determined by the interplay of quantum stochasticity due to impurity backscattering and the Fermi statistics~\cite{Blanter2000}.

Surprisingly enough, it has been predicted~\cite{Tworzydlo2006} that the same $F=1/3$ should be measured in {\em clean} wide and short graphene strips ($W/L \gtrsim 4$ where $W$ is the width and $L$ is the length of the strip). This is in sharp contrast to ballistic  transport (with perfect transmission) in 2DEG channels where shot noise is absent ($F=0$) due to completely correlated propagation of electrons by the Pauli principle. Such ``pseudo-diffusive''~\cite{Tworzydlo2006} conduction at the Dirac point  of graphene can be traced to metal-induced gap states~\cite{Golizadeh-Mojarad2007b}, familiar from metal-semiconductor junctions where they provide direct tunneling between evanescent scattering states of external leads populating the gap of a short sample. However, in graphene, viewed as a gapless semiconductor where valence and conduction bands touch at the Dirac point ($E_F=0$)~\cite{Geim2007}, the corresponding evanescent states~\cite{Robinson2007} can penetrate much longer distance
carrying current whose $F=1/3$ is accidentally~\cite{Tworzydlo2006} the same as in the diffusive metallic wires. In fact, $F=1/3$ has been  observed in recent experiments on large aspect ratio  ($W/L \simeq 24$, $L \sim 200$ nm) ballistic two-terminal graphene devices~\cite{Danneau2008}.

We recall that $F=1/3$ for two-terminal disordered conductors is {\em universal}---it does not depend  on the impurity arrangement, band structure, and the shape of the sample~\cite{Blanter2000}. Much less is know about the universality of the ``pseudo-diffusive'' shot noise in ballistic graphene strips where typically GNRs with armchair edges~\cite{Tworzydlo2006,Schomerus2007,Lewenkopf2008} or infinite gated ZGNR setups~\cite{Cresti2007} have been the subject of theoretical analysis. The topology of the zigzag edge is rather special generating peculiar edge-localized quantum states~\cite{Fujita1996}. Moreover, their partially flat (within one-third of 1D Brillouin zone) subband generates a large peak in the density of states at the Fermi energy $E_F=0$. This instability~\cite{Fazekas1999} is most likely resolved~\cite{Pisani2007} through magnetic ordering around the zigzag edge when electron-electron interactions (even infinitesimally small~\cite{Fujita1996}) are ``turned on.'' Such carbon-based  magnetism involving {\em s-p} orbitals was conjectured for an {\em infinite} ZGNR using the Hubbard model~\cite{Fujita1996}, and confirmed through numerous recent density functional theory (DFT) calculations~\cite{Son2006a,Pisani2007,Yazyev2008} and related proposals for spintronic devices~\cite{Son2006,Wimmer2008}.

\begin{figure}
\centerline{\psfig{file=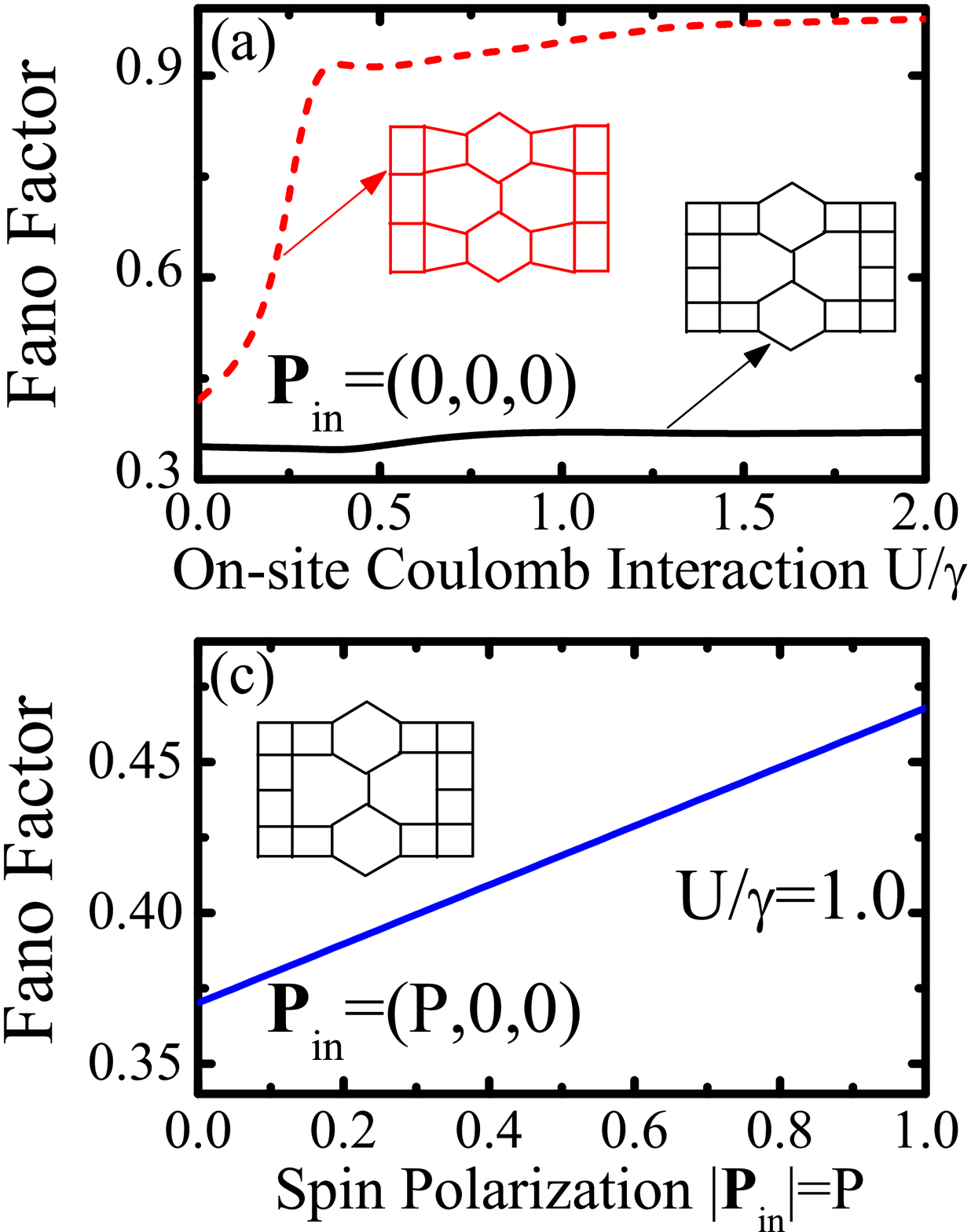,scale=0.15,angle=0}  \psfig{file=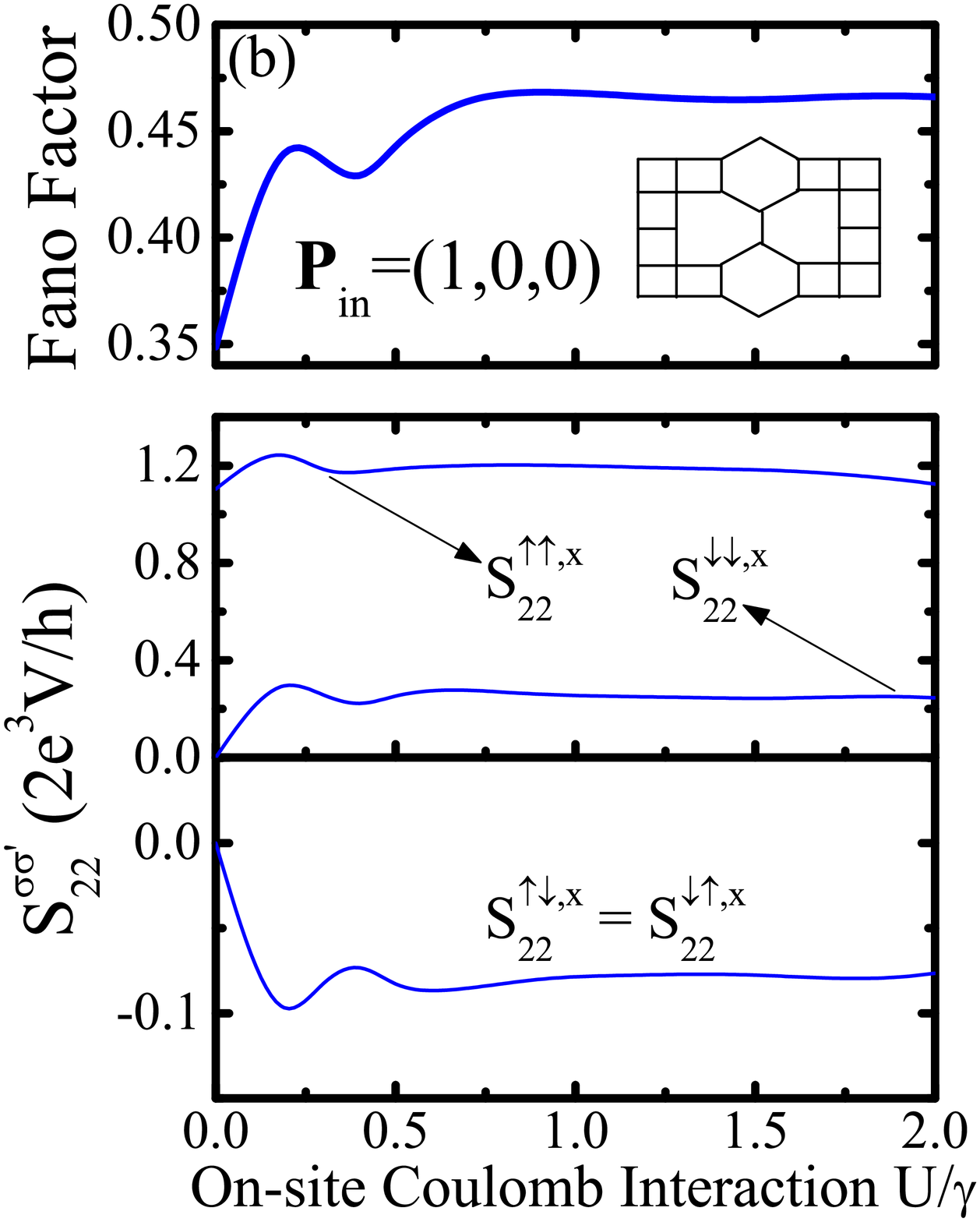,scale=0.15,angle=0}}
\caption{(Color online) The Fano factor of magnetically ordered  two-terminal 212-ZGNR  of length $N_a^z=75$ ($W/L=5$). Panel (a) is for unpolarized injection and two different types of metallic electrodes, whose square lattice is ``matched'' (solid line) or ``unmatched'' (dashed line) to the honeycomb lattice. Panel (b) plots total and spin-resolved shot noise for 100\% polarized injection, where injected spins are aligned with the transport direction (the $x$-axis) and both spins are detected in electrode 2. The Fano factor for partially polarized  injected current in the same setup as (b) is shown in (c).}
\label{fig:fano_int}
\end{figure}

Here we address these unresolved issues by predicting two experimentally testable effects in clean two-terminal
metal-ZGNR-metal devices that strongly intertwine their {\em magnetic} correlations with the {\em electrical} shot noise: ({\em i}) finite length ZGNR attached to metallic electrodes will also develop magnetic ordering around the edges which, however, decays in the vicinity of metallic
contacts and requires finite strength of Coulomb interaction (Fig.~\ref{fig:spin_density}); ({\em ii}) the magnetic moment per carbon
atom in the middle of ferromagnetically ordered zigzag edge is in one-to-one correspondence with the enhanced shot noise Fano factor $F > 1/3$, so that {\em all-electrical} measurements can be used as {\em edge-sensitive} technique to probe unusual {\em s-p} magnetism  in wide ZGNRs at
low temperatures (Fig.~\ref{fig:fano_int}).

We employ the single $\pi$-orbital Hubbard model~\cite{Fujita1996} to obtain magnetic ordering within ZGNR:
\begin{equation}\label{eq:hubbard}
\hat{H} = -\gamma \sum_{\langle \bf{i j} \rangle} \sum_{\sigma=\uparrow,\downarrow} (\hat{c}_{\bf{i}\sigma}^\dagger \hat{c}_{\bf{j}\sigma} + {\rm H.c.}) + U \sum_{\bf i} \hat{n}_{\bf i \uparrow} \hat{n}_{\bf i \downarrow},
\end{equation}
which is defined on a finite-size honeycomb lattice with zigzag edges and the lattice constant \mbox{$a=2.46$ \AA}.  Here $\hat{c}_{\bf i}^\dag$ ($\hat{c}_{\bf i}$) creates (annihilates) an electron in the $\pi$-orbital located at site ${\bf i}=(i_x,i_y)$ and $\gamma$ is the nearest neighbor hopping. The width of the $N_z$-ZGNR lattice is measured using the number $N_z$ of zigzag longitudinal chains~\cite{Fujita1996,Son2006a}. We use the number of atoms $N_a^z$ comprising a zigzag chain to  measure its length [in the units of $a$, the average width of ZGNR is $W=a\sqrt{3}(N_z-1)/2$ and its length is $L=(N_a^z-1)a/2$]. The metallic leads attached to ZGNR  are modeled by the tight-binding Hamiltonian ($U=0$) on the semi-infinite square lattice~\cite{Robinson2007,Schomerus2007}, which allows us to capture different contact effects~\cite{Robinson2007} introduced in experimental circuits~\cite{Geim2007,Danneau2008,DiCarlo2008} by ultimate electrodes being metals rather than all-graphitic structures.

\begin{figure}
\centerline{\psfig{file=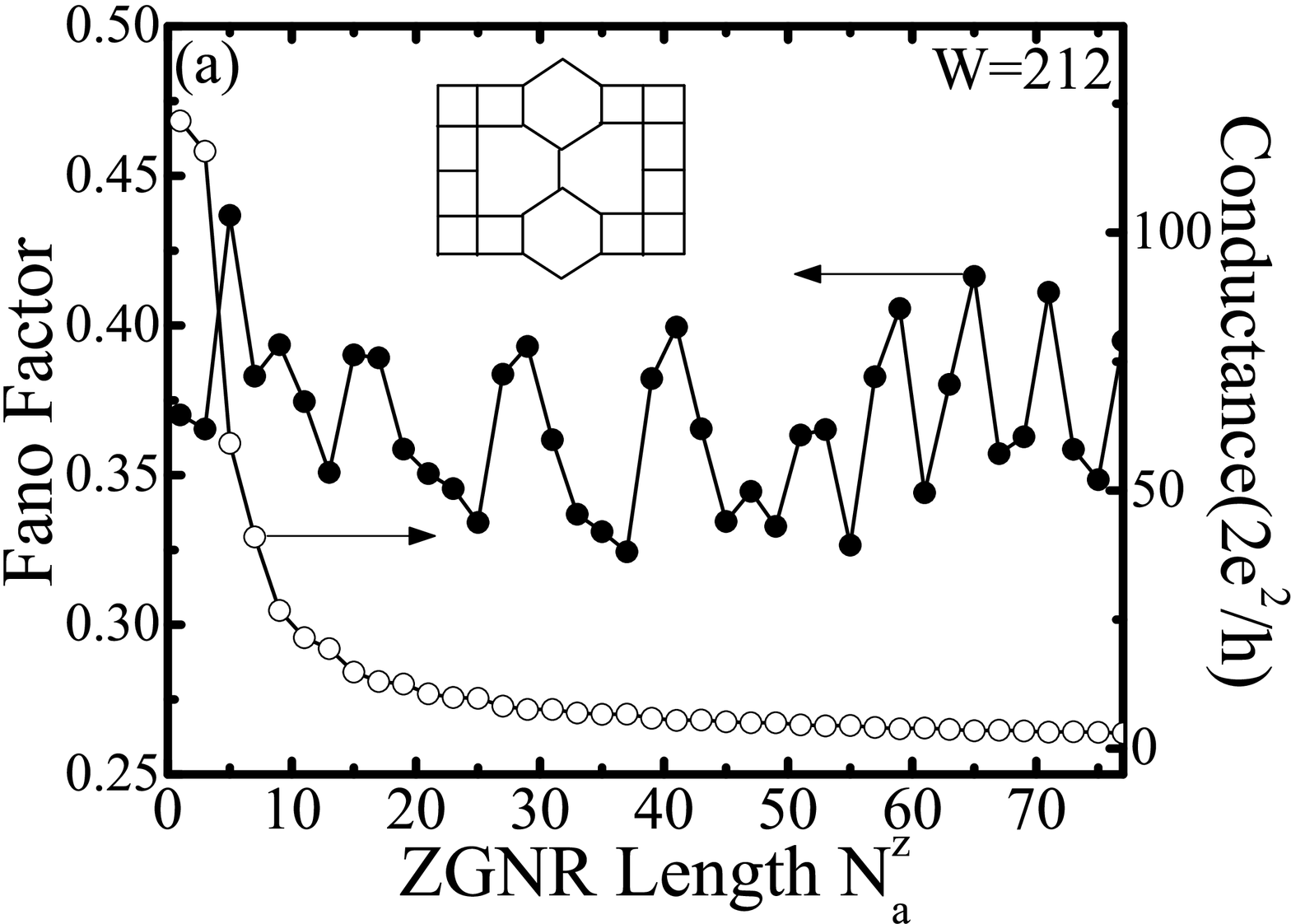,scale=0.20,angle=0} \hspace{0.05in} \psfig{file=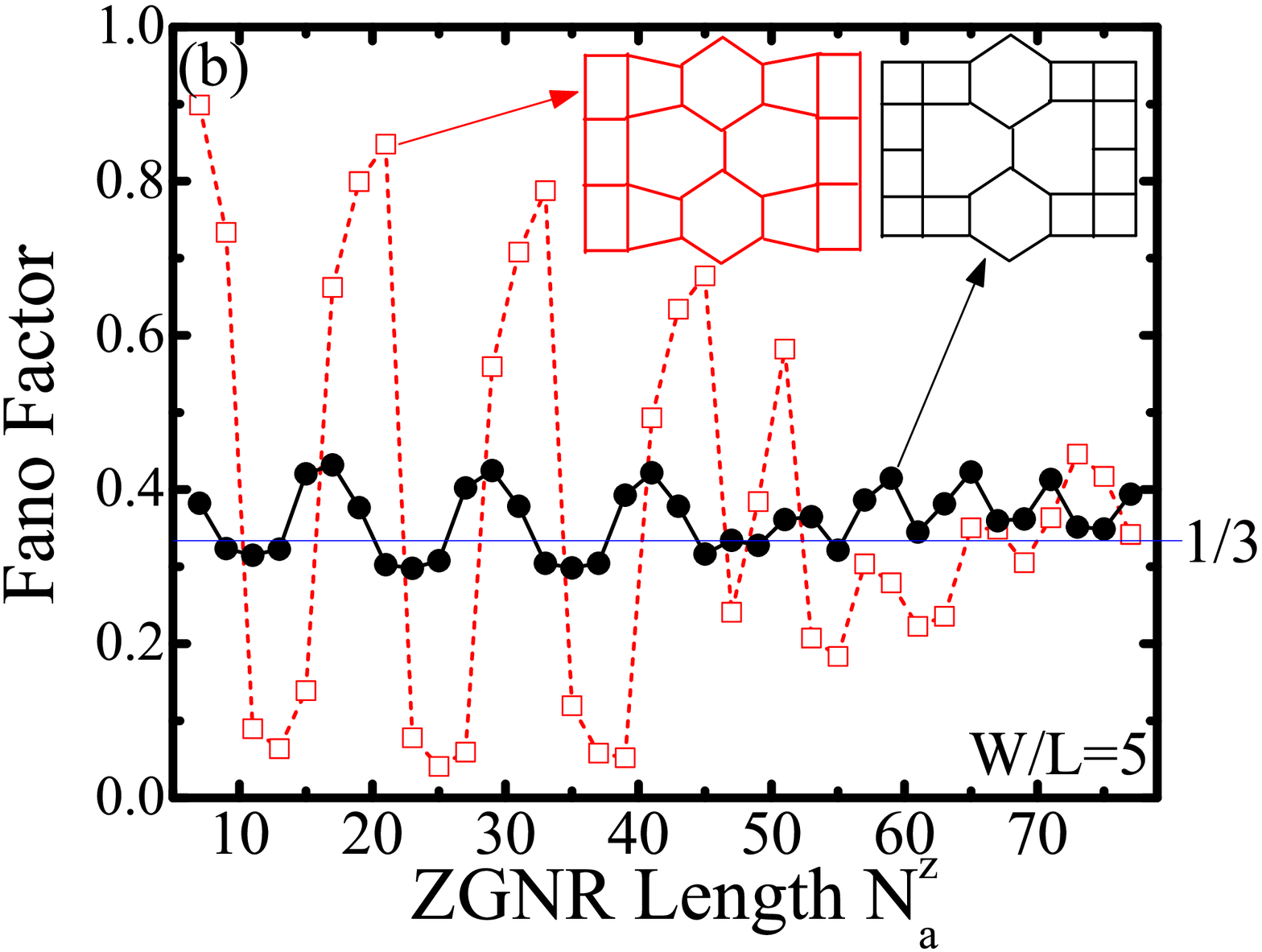,scale=0.20,angle=0}}
\caption{(Color online) (a) The Fano factor and conductance of nonmagnetic ($U=0$) 212-ZGNR attached to two metallic electrodes
modeled as square lattices. (b) Shot noise in ZGNRs of variable width $W$ and length $L$, but with their ratio fixed at $W/L=5$, for two different types of metallic electrodes whose square lattice is ``matched'' (solid line) or ``unmatched'' (dashed line) to the honeycomb lattice. The electrons are injected at the Fermi energy $E_F=10^{-6}\gamma$.}
\label{fig:fano_noint}
\end{figure}

\begin{figure}
\centerline{\psfig{file=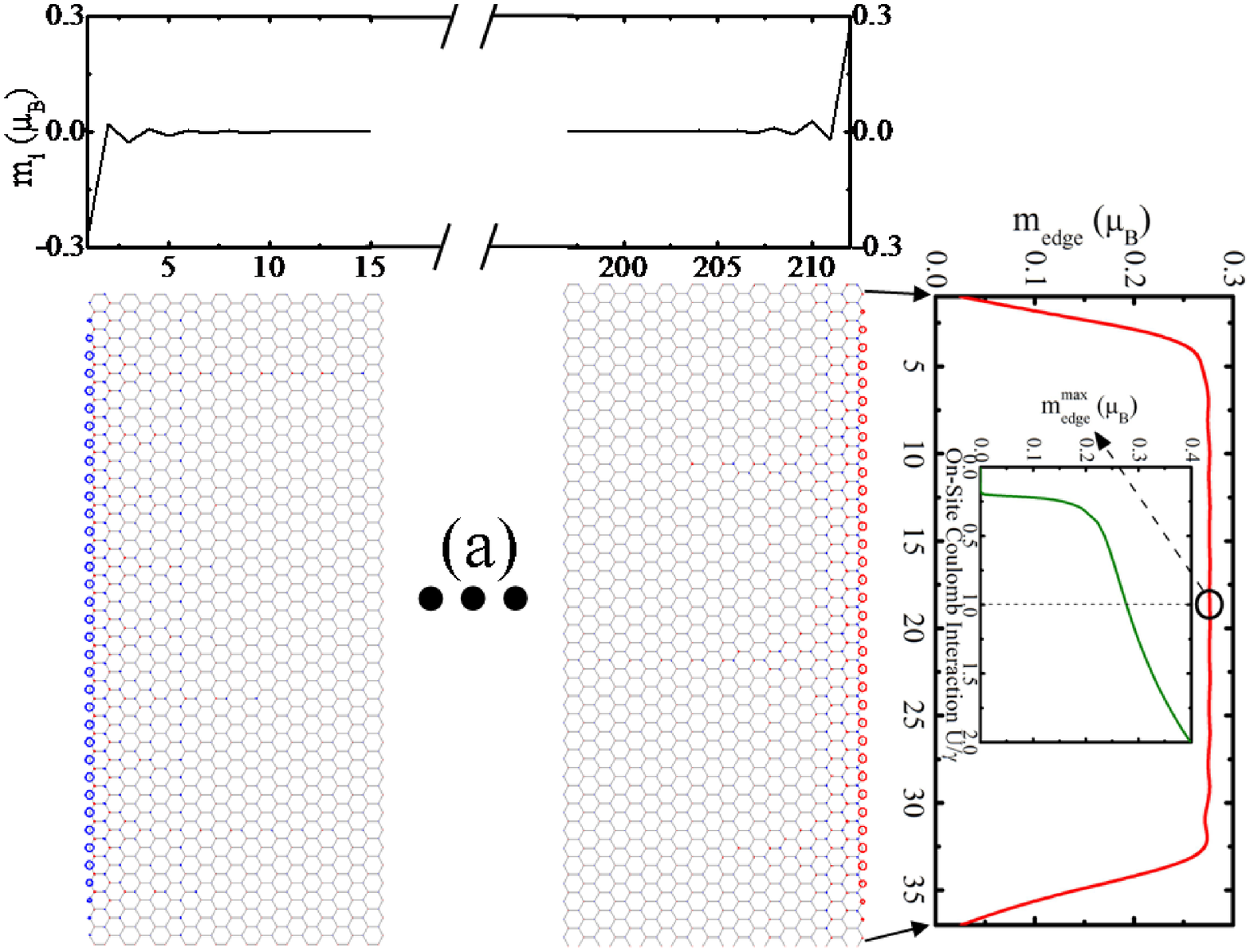,scale=0.28,angle=0}}
\vspace{0.05in}
\centerline{\psfig{file=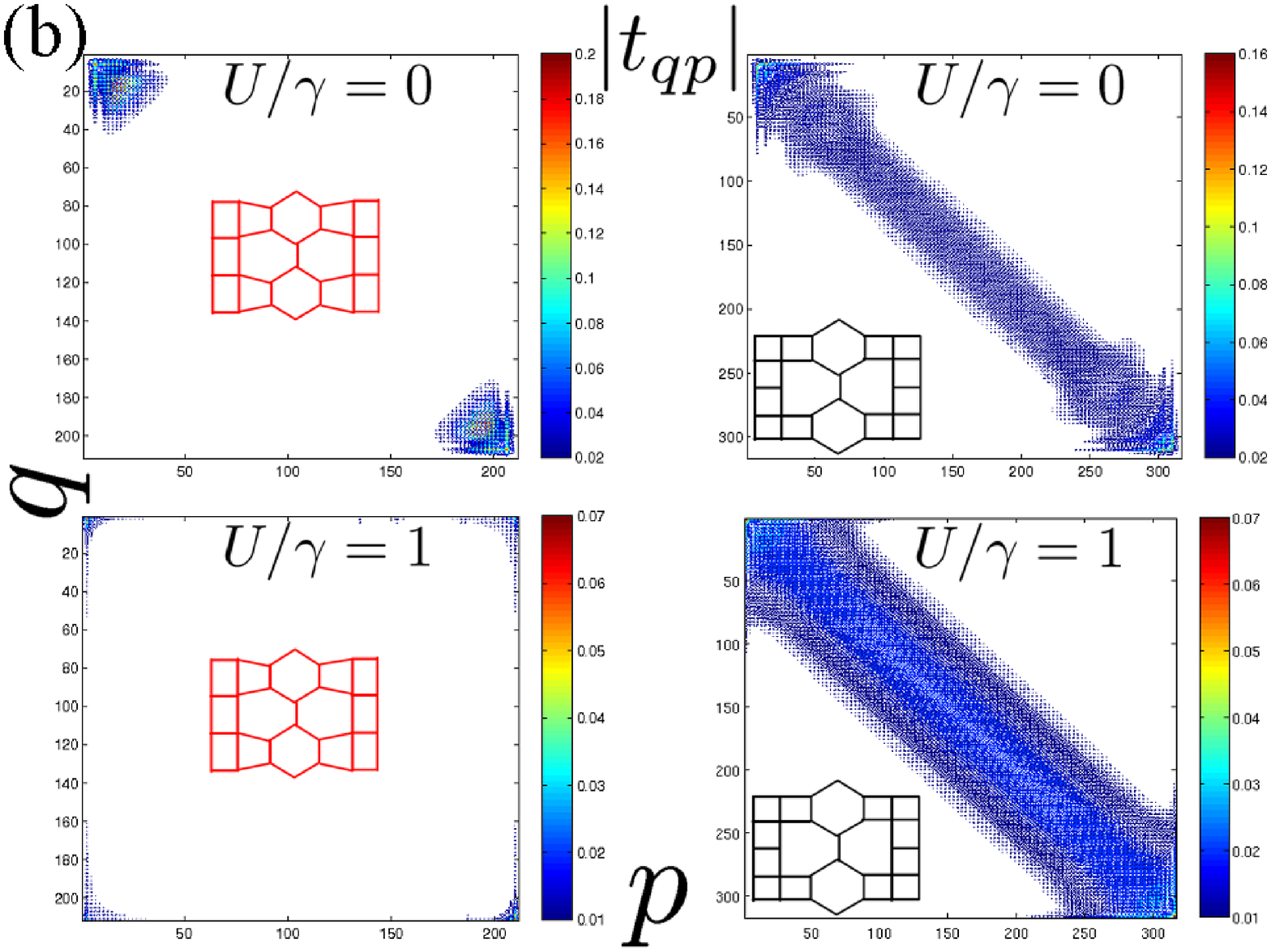,scale=0.22,angle=0}}
\caption{(Color online) (a) The spatial profile of the local magnetization $m_{\bf i}$ (pointing out-of-plane) within 212-ZGNR of length $N_a^z=75$ ($W/L=5$) induced by on-site Coulomb interaction $U/\gamma=1$. The metallic leads are modeled as semi-infinite square lattice attached at the top and bottom armchair interface. (b) Amplitude of the transmission matrix elements $|t_{qp}|$ for magnetically ($U/\gamma=1$ panels) ordered 212-ZGNR from panel (a) and its nonmagnetic ($U=0$ panels) version studied in Fig.~\ref{fig:fano_noint}, using ``lattice-unmatched''  (left column) or ``lattice-matched'' (right column) square lattice leads.}
\label{fig:spin_density}
\end{figure}

By setting $U=0$ both in the electrodes and in the ZGNR central region, the Fano factor in Fig.~\ref{fig:fano_noint}(a) is obtained using the celebrated scattering theory formula~\cite{Blanter2000}, $F=\sum_n T_n(1-T_n)/\sum_n T_n$. Here $T_n$ are the eigenvalues of ${\bf t} {\bf t}^\dag$ and ${\bf t}$ is the transmission matrix of a phase-coherent device. In the basis of eigenchannels that diagonalize ${\bf t} {\bf t}^\dag$, a mesoscopic device can be viewed as a parallel circuit of independent one-dimensional conductors. Figure~\ref{fig:fano_noint}(b) shows that Fano factor is substantially affected by the type of the square lattice leads~\cite{Robinson2007}, even in samples with large $W/L$. They effectively introduce disorder at the lead/ZGNR interface, thereby mixing the transverse propagating modes. That is, ${\bf t}$ acquires non-zero off-diagonal elements even in clean devices, as shown in Fig.~\ref{fig:spin_density}(b). The absence of mode mixing in transport through armchair GNR,  attached to highly doped graphene~\cite{Tworzydlo2006} or square lattice metallic leads~\cite{Schomerus2007,Lewenkopf2008,Robinson2007}, is intimately connected to its $F=1/3$ value~\cite{Tworzydlo2006,Schomerus2007,Lewenkopf2008}. Nevertheless, for metallic leads in Fig.~\ref{fig:fano_noint}(a), whose lattice matches~\cite{Robinson2007} the honeycomb lattice (square lattice spacing is the same as the carbon-carbon distance) while allowing propagating modes in the leads to efficiently couple via evanescent ones in ZGNR in Fig.~\ref{fig:spin_density}(b), we find Ohmic-like conductance $G=\frac{2e^2}{h}\sum_n T_n \sim W/L$ in Fig.~\ref{fig:fano_noint}(a) and the Fano factor $F \simeq 0.3 - 0.4$ for $W/L=5$ in Fig.~\ref{fig:fano_noint}(b).

We select 212-ZGNR of length $N_a^z=75$, whose dimensions ensure $F \simeq 1/3$ in Fig.~\ref{fig:fano_noint}(b), for the analysis of interaction $U > 0$ driven magnetism and its effects on the shot noise. In infinite  ZGNRs, edge states manifest as two peaks in the local density of states along the edge~\cite{Zarbo2007}. The overlap of states from two edges yields bonding and anti-bonding states enabling a single conducting channel close to the Dirac point $E_F=0$ with highly unusual transport properties~\cite{Zarbo2007}. The evanescent state enabled Ohmic-like transport $G \propto W/L$ in finite ZGNR devices is expected at energies $E_F < \Delta E$ of the single open conducting channel~\cite{Cresti2007}. Within non-interacting ($U=0$)  212-ZGNR this channel is the only open one when $E_F$ of the injected electrons is $E_F < \Delta E = 0.015 t$~\cite{Zarbo2007}, while 317 channels are used for injection from the square lattice lead in Fig.~\ref{fig:fano_noint}(a) at half filling. The Fano factor and conductance of this ZGNR are determined by few non-negligible transmission eigenvalues, out of which one is close to unity corresponding to transport through the edge states induced channel.

To capture the magnetic ordering in wide ZGNR two-terminal device, we use the standard mean-field decoupling scheme~\cite{Fazekas1999} which yields an effective~\cite{Fujita1996,Wimmer2008}  single-particle  Hamiltonian
\begin{equation}\label{eq:mf}
\hat{H}_{\rm MFA}=\sum_{{\bf i},\sigma} \frac{\Delta_{\bf i}}{2} \hat{c}_{\bf{i}\sigma}^\dagger \hat{\sigma}^z_{\sigma\sigma}  \hat{c}_{\bf{i}\sigma} -t \sum_{\langle \bf{i j} \rangle,\sigma} (\hat{c}_{\bf{i}\sigma}^\dagger \hat{c}_{\bf{j}\sigma} + {\rm H.c.}).
\end{equation}
 The inhomogeneous exchange potential  $\Delta_{\bf i} = - U (\langle \hat{n}_{\bf i \uparrow} \rangle - \langle \hat{n}_{\bf i \downarrow} \rangle)$ determines the spin splitting, where $\hat{\sigma}^z_{\sigma\sigma}$ are elements of the Pauli matrix. The local magnetic ordering is quantified by the $z$-component of spin  $\langle \hat{S}^z_{\bf i} \rangle = (\langle \hat{n}_{\bf i \uparrow}\rangle - \langle \hat{n}_{\bf i \downarrow}\rangle)/2$ and the local magnetization $m_{\bf i}=g \mu_B \langle \hat{S}^z_{\bf i} \rangle$. The average electron density in equilibrium
\begin{subequations}\label{eq:charge}
\begin{eqnarray}
\langle n_{\bf i \sigma} \rangle & = & -\frac{1}{\pi} \! \int\limits_{-\infty}^{+\infty} \!\!dE \, {\rm Im}\, \langle {\bf i} \sigma |\hat{G}^r(E)|{\bf i} \sigma \rangle f(E-E_F), \\
\hat{G}^r(E) &  = & [E-\hat{H}_{\rm MFA} -\hat{\Sigma}_1 -\hat{\Sigma}_2]^{-1},
\end{eqnarray}
\end{subequations}
is computed from the diagonal matrix elements of the retarded Green operator $\hat{G}^r(E)$ for the open system ZGNR+leads~\cite{Areshkin2009}. The retarded self-energies operators $\hat{\Sigma}_1$, $\hat{\Sigma}_2$ introduced by the interaction with the leads determine escape rates of electrons into the electrodes.

Equations~(\ref{eq:mf}) and (\ref{eq:charge}) are solved self-consistently with charge convergence accelerated by using small non-zero temperature in the Fermi function $f(E-E_F)$ describing the electrodes~\cite{Areshkin2009}. The result for the local magnetization plotted in Fig.~\ref{fig:spin_density}(a) shows usual staggered pattern with salient feature being ferromagnetically ordered spins along each edge and opposite spin directions between the edges as the {\em ground} state~\cite{Fujita1996,Son2006a,Pisani2007}. However, Fig.~\ref{fig:spin_density}(a) highlights two major differences between finite ZGNRs sandwiched between two metallic leads and infinite ZGNR: ({\em i}) the magnetic moment per edge carbon atom decays upon approaching the metallic lead, so that no magnetic ordering appears in ZGNR shorter than $N_a^z \simeq 6$; ({\em ii}) while infinitesimally small $U$ causes ferrimagnetic ordering on each sublattice around the edges of an infinite ZGNR~\cite{Fujita1996}, in two-terminal ZGNR devices $U/\gamma \gtrsim 0.2$ is required to get sizable $m_{\rm edge}$ on  outermost carbon atoms, as shown in the inset of  Fig.~\ref{fig:spin_density}(a).

Although DFT  calculations go beyond only on-site Coulomb interaction and nearest-neighbor hopping of $\hat{H}_{\rm MFA}$, they can be mapped~\cite{Pisani2007,Areshkin2009} to electronic and magnetic structure at half-filling (we use $E_F=10^{-6}\gamma$) obtained from simpler Eq.~(\ref{eq:mf}). The replacement of $\hat{H}_{\rm DFT}$ with minimal-basis-set $\hat{H}_{\rm MFA}$ allows us to treat systems composed of $\sim 10^5$ carbon atoms. The values for $U$ and $\gamma$ extracted via this mapping depend slightly on the type of approximation scheme for exchange-correlation density functional~\cite{Pisani2007}. We find that $U/\gamma =1$ in Fig.~\ref{fig:spin_density}(a) reproduces the {\em ab initio}  result~\cite{Yazyev2008} for the magnetic moment $m_{\rm edge}^{\rm max} \approx 0.28$ $\mu_B$ per carbon atom in the middle of a zigzag edge, as well as the energy gap $\Delta_z^0$~\cite{Areshkin2009}  of an infinitely long ZGNR opening around $E_F=0$ due to the staggered sublattice potential~\cite{Son2006a}. The gap  $\Delta_z^0$ is experimentally visible~\cite{Li2008} in very narrow ribbons and vanishes~\cite{Pisani2007} within the room-temperature thermal energy window when the width of ZGNR reaches $\simeq 80$ nm.

The self-consistently computed effective Hamiltonian Eq.~(\ref{eq:mf}) is unaffected by small voltage bias~\cite{Areshkin2009} and can be used as the basis for linear response transport calculations within single-particle formalisms~\cite{Wimmer2008}. While recent studies have employed such Hamiltonian to obtain the spin-resolved conductances of two-terminal ZGNR devices~\cite{Wimmer2008}, here we utilize the scattering approach to quantum transport~\cite{Blanter2000} to compute the spin-dependent shot noise as quantified by correlators between spin-resolved charge currents $I^\uparrow_2$ and $I^\downarrow_2$ in the drain lead 2:
\begin{equation}\label{eq:noise}
S_{22}^{\sigma \sigma^\prime} (t-t^\prime) = \frac{1}{2} \langle \delta \hat{I}_2^\sigma(t) \delta \hat{I}_2^{\sigma^\prime}(t^\prime) +  \delta \hat{I}_2^{\sigma^\prime}(t^\prime)  \delta \hat{I}_2^\sigma(t) \rangle.
\end{equation}
Here $\hat{I}_2^\sigma(t)$ is the quantum-mechanical operator of spin-resolved  charge current of spin-$\sigma$ electrons in lead $2$. The current-fluctuation operator at time $t$ in lead $2$ is $\delta \hat{I}_2^\sigma(t) = \hat{I}_2^\sigma(t) - \langle \hat{I}_2^\sigma (t) \rangle$ and $\langle \ldots \rangle$ denotes both quantum-mechanical and statistical averaging~\cite{Blanter2000} over the states in the macroscopic reservoirs to which ZGNR conductor is attached via semi-infinite interaction-free metallic leads. The Fourier transform $S_{22}^{\sigma \sigma^\prime} (\omega) = 2 \int d(t-t^\prime)\, e^{-i\omega(t-t^\prime)} S_{22}^{\sigma \sigma^\prime} (t-t^\prime)$ gives the spin-resolved noise power, and the total charge current  noise is obtained from $S_{22}^{\rm ch}(\omega) = \sum_{\sigma,\sigma'} S_{22}^{\sigma\sigma^\prime}(\omega)$ where we focus on the  zero-frequency limit $S_{22}^{\rm ch}=S_{22}^{\rm ch}(\omega \rightarrow 0)$ of these expressions.

The presence of intrinsic (due to exchange interaction, spin-orbit couplings, magnetic impurities, etc.) and external magnetic fields in the sample is manifested in the shot noise dramatically when injected current is spin-polarized~\cite{Dragomirova2007,Nikoli'c2009,Hatami2006}. This is due to the fact that any spin flip converts spin-$\uparrow$ subsystem particle into a spin-$\downarrow$ subsystem particle, where the two subsystems differ when spin degeneracy if lifted. Thus, the nonconservation of the number of particles in each subsystem generates additional source of current fluctuations. In particular, the shot noise of spin-polarized current injected from ferromagnetic reservoirs into a ferromagnetic wire was predicted to be a sensitive probe of its magnetic ordering and spin-dependent interactions~\cite{Hatami2006}.

The computation of $S_{22}^{\rm ch}$ for the general situation, when injected spins are characterized by a density matrix $\hat{\rho}^s_{\rm in}$, cannot be done using the standard basis of eigenchannels and the corresponding transmission eigenvalues $T_n$ since in this approach the information about $\hat{\rho}^s_{\rm in}$ is lost. Instead, one has to use the lengthy formulas, derived in Ref.~\cite{Dragomirova2007} in terms of the spin-resolved transmission matrix  ${\bf t}_{2 1}^{\sigma \sigma^\prime}  =   2 \sqrt{-\text{Im} \, \hat{\Sigma}_2^{r,\sigma}} \cdot \hat{G}_{2 1}^{r,\sigma \sigma^\prime} \cdot \sqrt{-\text{Im}\, \hat{\Sigma}_1^{r,\sigma'}}$ of the device, to keep track of $\hat{\rho}^s_{\rm in}=(1+{\bf P}_{\rm in} \cdot \hat{\bm \sigma})/2$ and its spin polarization vector ${\bf P}_{\rm in}$. Although no true long-range order is possible in one dimension (the spin correlation length along the zigzag edge decays to $\sim 1$ nm at room temperature~\cite{Yazyev2008}), the observation of the shot noise requires low temperature~\cite{Danneau2008} where magnetic correlations should be highly visible. Therefore, we use $T=0$ scattering formulas~\cite{Dragomirova2007,Nikoli'c2009} for $S_{22}^{\sigma \sigma^\prime}$. The electrode 2 is paramagnetic so that it collects  both spin species.

The Fano factor $F({\bf P}_{\rm in}) = S_{22}^{\rm ch}/2 e I_2$ for unpolarized $|{\bf P}_{\rm in}|=0$ current injection from electrode 1 into ZGNR is shown in Fig.~\ref{fig:fano_int}(a). In the case of ``lattice-matched'' leads, the sensitivity of transport around $E_F=0$ to edge potential [and reflection from it, Fig.~\ref{fig:spin_density}(b)] drives  slight increase ($F \simeq 0.37$ at $U/\gamma=1$) of the Fano factor above the reference value $F \simeq 0.35$ at $U=0$. The enhancement---$F \simeq 0.95$ at $U/\gamma=1$---is much more dramatic for ``lattice-unmatched'' leads, where Fig.~\ref{fig:spin_density}(b) shows that even at $U=0$ propagating modes in two leads are decoupled due to poor matching (the conductance does not scale as $\sim  W/L$)  to most slowly~\cite{Robinson2007} decaying evanescent modes. At $U>0$, a small gap ($\Delta_z^0 \simeq 0.009\gamma$ at $U/\gamma=1$) opens in 212-ZGNR around $E_F=0$ which drastically reduces the transmission matrix elements [Fig.~\ref{fig:spin_density}(b)], thereby, converting transport into conventional tunneling with Fano factor close to $F=1$.

For injected spins collinear with the local magnetic moments within ZGNR there is no spin precession and cross-correlators are zero $S_{22}^{\uparrow \downarrow,z} \equiv 0$. The shot noise of spin-polarized current becomes non-trivial when injected spins are non-collinear with ${\bf m}_{\bf i}$, as is the case of the $x$-axis (direction of transport) polarized spins in Fig.~\ref{fig:fano_int}(b). Since they are not the eigenstates of the local magnetic field along the $z$-axis, they are forced into precession~\cite{Nikoli'c2009}. This enhances auto-correlation  noise $S_{22}^{\uparrow \uparrow,x}$ and generates non-zero cross-correlators $S_{22}^{\uparrow \downarrow,x} < 0$. Figure~\ref{fig:fano_int}(c) suggests that probing of ZGNR edge magnetic ordering via spin-dependent shot noise can be efficient even with partially polarized $0<|{\bf P}_{\rm in}|<1$ injected current.

{\em Conclusions}---Our first principal result describes magnetic ordering in wide and short  ZGNRs where magnetization persists at the edge (with no interedge magnetic order characterizing narrow ZGNRs) even as its width goes to infinity. The edge magnetization diminishes in the longitudinal direction upon approaching the   \mbox{ZGNR$|$metallic-electrode} interface. This picture complements recent DFT calculations~\cite{Pisani2007,Son2006a,Yazyev2008} on narrow infinite ideal ZGNRs. Since in the limit of wide ZGNR carbon-based {\em s-p} magnetism becomes a pure edge effect---not detectable by bulk-sensitive techniques---we offer a recipe on how spin-dependent shot noise can be exploited as {\em edge-sensitive} and {\em all-electrical} probe of this phenomenon. That is, Fig.~\ref{fig:fano_int} suggests an experiment that would: inject unpolarized current into ballistic ZGNR with ultrasmooth edges~\cite{Li2008} $\Rightarrow$ measure possible enhancement of the shot noise above recently observed $F=1/3$~\cite{Danneau2008} $\Rightarrow$ inject in-plane polarized current while collecting both spins to observe further Fano factor enhancement with increasing spin polarization $|{\bf P}_{\rm in}|$ due to precession of transported spins in edge magnetic field.

\begin{acknowledgments}
Financial support from NSF Grant No. ECCS 0725566 and DOE Grant No. DE-FG02-07ER46374 is gratefully acknowledged.
\end{acknowledgments}



\end{document}